\documentclass[pra,onecolumn,amsmath,amssymb]{revtex4}

\usepackage{bm,epic,eepic}
 \usepackage[english]{babel}
 \usepackage{graphicx}
 \usepackage{pst-plot}
  \usepackage{pstricks}
   \usepackage{pst-coil}

\def\renorm{\ensuremath{\frac{1}{\sqrt{2}}}}
\def\half{\ensuremath{\frac{1}{2}}}

\def\suk#1{\ket{{#1}\uparrow}}
\def\sdk#1{\ket{{#1}\downarrow}}

\def\bra#1{\ensuremath{\mathinner{\langle{#1}|}} }
\def\ket#1{\ensuremath{\mathinner{|{#1}\rangle}} }

\begin{document}

\title{A local hidden-variable model violating Bell's inequalities: a reply to Matzkin.}
\date{\today}

\author{R.N. Nijhoff}
\email{reinder@infi.nl} \affiliation{Institute of History and
Foundations of Science, Utrecht University, The Netherlands}

\begin{abstract} 
Recently, Matzkin claimed the construction of a hidden variable (HV) model \cite{Matzkin-2007}  which is both local and equivalent with the quantum-mechanical predictions. 
In this paper we will briefly present this HV model and argue, by identifying an extra non-local ``hidden'' HV , why this model is not local.
\end{abstract} 

\maketitle

\section{Introduction}

With the derivation of his well-known inequalities, Bell \cite{Bell} proved that any local model based on hidden variables (HV) can not reproduce the empirical predictions of quantum-mechanics. Recently however, Matzkin claimed that he has constructed a HV model \cite{Matzkin-2007}  which is both local and equivalent with the quantum-mechanical predictions. 

Such a claim is remarkable, especially when we take into account that stochastic local HV models (like the HV model of Matzkin) also have to obey the Bell inequalities \cite{Bell2}. In this paper we will briefly present the HV model of Matzkin and argue why this model is not local.

\section{EPRB}

In the EPRB-experiment, studied by both Bell and Matzkin,  a pair of spin-$\frac{1}{2}$ particles are formed in the singlet spin state $\ket{\Psi_0}$:
\begin{equation}
\ket{\Psi_0} = \renorm( \suk{z} \otimes \sdk{z} - \sdk{z} \otimes \suk{z}).
\end{equation}
Both particles move freely in opposite directions towards two measuring devices. The spin component $a$ of particle 1 is measured in direction $\vec{a}$ and the spin component $b$ of  particle 2 is measured in direction $\vec{b}$. If we measure the spin component for both particles in the same direction ($\vec{a} = \vec{b}$), anti-correlation of the particles leads to the measurement outcome $a = -b$. More generally, for arbitrary directions $\vec{a}, \vec{b}$, the quantum-mechanical expectation value of the product of the two spin components $<S_{\vec{a}}  \otimes S_{\vec{b}}>$ is:
\begin{equation}\label{eq_eqm}
E_{QM} ( \vec{a},\vec{b} ) \equiv \bra{\Psi_0} S_{\vec{a}} \otimes S_{\vec{b}} \ket{\Psi_0} = - \frac{ \vec{a} \cdot \vec{b}  }{4},
\end{equation}
where $S_{\vec{a}} = \frac{1}{2}(\vec{\sigma}\cdot \vec{a})$ and $S_{\vec{b}} = \frac{1}{2}(\vec{\sigma}\cdot \vec{b})$.

\section{Matzkins HV model}

Matzkins HV model makes use of the following elements:

\begin{enumerate}
\item \emph{HV.} A single particle is specified by a hidden variable $\vec{\lambda}$: a normalized vector in $\mathbb{R}^3$. 
For the two particles specified by $(\vec{\lambda}_1, \vec{\lambda}_2)$  in the singlet spin state, 
anti-correlation is described by the relation:
\begin{equation}\label{eq_anti}
\vec{\lambda}_1 = -\vec{\lambda}_2.
\end{equation}
\item \emph{HV distribution.} 
As we will see, the measurement outcome of a measurement on a single particle does not only depend on the HV $\vec{\lambda}$ describing the particle, but it does also depend on a HV distribution $R$.
Matzkin states that a one-particle system in an eigenstate of the spin-operator in (the normalized) direction $\vec{u}$ with a spin component $u = \pm \half$, can be described with a HV distribution:
\begin{equation}\label{eq_dist_sp}
R_{\pm \half \vec{u}}(\vec{\lambda}) = \frac{1}{\sqrt{3}\pi} \delta(\vec{\lambda} \cdot \vec{u} \mp \half).
\end{equation}
In his paper, Matzkin does not make clear how we have to interpret this HV distribution $R$.
The fact that two particles in the singlet state, are anti-correlated by relation \eqref{eq_anti} however, suggest that a single particle is characterized by only one HV $\vec{\lambda}$.
It is therefore plausible that the distribution $R$ is a probability distribution which describes an ensemble of all different $\vec{\lambda}$ corresponding to the same quantum state of the system.
Each element of the ensemble corresponds thereby with a single particle described by one HV $\vec{\lambda}$.

\item \emph{Measurement outcome.} A measurement induces a perturbation of the initial HV distribution. 
If we measure the spin component of a particle in direction $\vec{u}$, the perturbation of the \emph{initial} HV distribution ($R^{t_0}$) depends on the measurement outcome $\mathcal{M}(R^{t_0}, \vec{u},\vec{\lambda}) = \pm \half$. 
Matzkin states that for such a spin measurement in direction $\vec{u}$, the \emph{post-measurement} HV distribution ($R^{t_1}$) will be equal to the distributions given in \eqref{eq_dist_sp}, depending on the measurement outcomes $\pm \half$. 

For a system consisting of a single particle in a quantum state with a positive spin along the $\vec{z}$-axis (having an initial HV distribution $R_{\pm \half \vec{z}}$) the probabilities of $\mathcal{M}(R=R_{+\half\vec{z}},\vec{u},\vec{\lambda}) = \pm \half$ by measuring the spin component in direction $\vec{u}$ are given by and equal to the quantum-mechanical probabilities:
\begin{eqnarray}\label{eq_ch_qm}
P(\mathcal{M}(R=R_{+\half\vec{z}},\vec{u},\vec{\lambda}) = \half ) = \cos^2 \frac{\theta_{\vec{z},\vec{u}}}{2} \\
P(\mathcal{M}(R=R_{+\half\vec{z}},\vec{u},\vec{\lambda}) = -\half ) = \sin^2 \frac{\theta_{\vec{z},\vec{u}}}{2}
\end{eqnarray}
with $\theta_{\vec{z},\vec{u}}$ the angle between $\vec{z}$ and $\vec{u}$.

Note that this probabilities not only depend on $\vec{\lambda}$ and $\vec{u}$, but are also dependent on the initial HV distribution $R$.
Because there is no correlation between $\vec{\lambda}$ and the HV distribution of a system, we make this dependency clear by adding $R$ as an extra (hidden) variable:
\begin{equation}
\mathcal{M}(R, \vec{u}, \vec{\lambda}) = \pm \half.
\end{equation}
Because it is important for our argument, we stress this point again: in general, the outcome of a measurment in de HV model of Matzkin does not only depend on a HV $\vec{\lambda}$, but it does also depend on a HV distribution $R$.
We therefore recognize this HV distribution $R$ as an extra hidden variable of the model.

\item \emph{Locality.} The only statement Matzkin makes about locality is that ``\ldots the HV distribution is dynamically and locally affected by the measurement \ldots''. 
To be more precise we assume that, following Bell, for a HV model the measurement outcome of particle 2 has to be a function of information locally available at particle 2. In particular, it may therefore not depend on the measurement direction and measurement outcome of a measurement on particle 1.
\end{enumerate}

\section{A HV model for two particles}

With these elements a HV-model for  the EPRB experiment is constructed. The system consists of two particles, described with the HV $(\vec{\lambda}_1, \vec{\lambda}_2)$ respecting correlation \eqref{eq_anti}.
The initial distribution of HV for each particle in the singlet state is (as given by Matzkin) a uniform distribution  $R_{\Sigma}$ of HV on a unit sphere. As noted, the measurement outcome of a measurement on one of the two particles depends on the measurement direction $\vec{u}$, the HV $\vec{\lambda}$ and the initial HV distribution. 
For a distribution $R_{\Sigma}$, the measurement outcome $\mathcal{M}$ of a measurement on one of the particles (described by $\vec{\lambda}$), is defined by:
\begin{equation}\label{eq_m}
\mathcal{M}(R=R_{\Sigma}, \vec{u}, \vec{\lambda}) = \half \frac{ \vec{u} \cdot \vec{\lambda}  } { ||\vec{u} \cdot \vec{\lambda} || } \in \{ -\half, \half \}
\end{equation}
Simply stated: the sign of the measurement outcome depends on the sign of the inner product of the HV $\vec{\lambda}$ and the measurement direction $\vec{u}$.

Knowing the initial HV distributions $R_\Sigma$ for the two particles in the singlet state, formula \eqref{eq_anti} describing the relation between $(\vec{\lambda}_1, \vec{\lambda_2})$  and formula \eqref{eq_m}, we are able to calculate the expectation value $E_{HV}(\vec{a},\vec{b})$ for the HV model of Matzkin. 
Integrating over the initial HV distribution and using relation \eqref{eq_anti} gives:
\begin{eqnarray}\label{eq_evvt1}
E_{HV}(\vec{a},\vec{b}) = <\mathcal{M}_{1}(R_\Sigma,\vec{a}), \mathcal{M}_{2}(R_\Sigma,\vec{b})> = \nonumber\\
\int 
\mathcal{M}_{1}(R_\Sigma,\vec{a}, \vec{\lambda}_1)
\mathcal{M}_{2}(R_\Sigma,\vec{b}, -\vec{\lambda}_1) \frac{1}{4\pi} d\Omega_{\vec{\lambda}_1} = -\frac{1}{4} + \frac{1}{2 \pi} \theta_{\vec{a},\vec{b}}.
\end{eqnarray}
Unfortunately this result is not equal to the quantum-mechanical prediction \eqref{eq_eqm}. In fact, the result is exactly the result of a naive HV model proposed by Bell \cite{Bell}, which does not violate the Bell inequalities at all.

Of course \eqref{eq_evvt1} is not the result of Matzkin. So what went wrong? 
The difference is that Matzkin makes an extra assumption before calculating the expectation value $E_{HV}$. 
Matzkin states that if we measure the spin component of particle 1 along axis $\vec{a}$ ($S_{\vec{a}}$) and obtain, for example, result $a = +\half$, we \emph{know} (using \eqref{eq_m}) that $\vec{\lambda}_1 \cdot \vec{a} \geq 0$. 
With the anti-correlation \eqref{eq_anti} in mind, we therefore also \emph{know} that $\vec{\lambda}_2 \cdot \vec{a} \leq 0$.
After a measurement of the spin of particle 1, we will therefore know the HV $\vec{\lambda}_2$ (describing particle 2) more precisely.

By knowing $\vec{\lambda}_2$ more precisely, Matzkin continues, we are also able to limit the initial HV distribution $R$ for particle 2.
Given the original uniform distribution $R_{\Sigma}$ and the relation $\vec{\lambda}_2 \cdot \vec{a} \leq 0$, the HV $\vec{\lambda}_2$ is, for example, also certainly an element of the uniform distribution on a half-sphere $R_{\Sigma_{-\half\vec{a}}}$.
We can therefore, Matzkin argues, use $R_{\Sigma_{-\half\vec{a}}}$ as the HV distribution of particle 2 after measuring $S_{1,\vec{a}} = +\half$.

By postulating that a particle described with a HV distribution $R_{\Sigma_{-\half\vec{a}}}$ behaves like it is in an eigenstate of $S_{\vec{a}}$ with spin component $a = -\half$, Matzkin argues that the measurement-outcomes $\mathcal{M}$ for such a particle will obey the (quantum-mechanical) probabilities given in \eqref{eq_ch_qm}. 
If so, the calculated expectation value $E_{HV}$ will correspond to the quantum-mechanical prediction \eqref{eq_eqm}\footnote{
Matzkin uses a different notation in his paper, making the dependency between the measurement direction and outcome of the measurement of the spin of particle 1 and the measurement outcome of a measurement on particle 2 less clear. He writes the distributions as a subscript and directly substitutes the probabilities of \eqref{eq_ch_qm} in the integral.
}:
\begin{eqnarray}\label{eq_evvt2}
E_{HV}(\vec{a},\vec{b}) = <\mathcal{M}_{1}(\vec{a}), \mathcal{M}_{2}(\vec{b})> = \nonumber\\
\int 
\mathcal{M}_{1}(R_\Sigma,\vec{a}, \vec{\lambda}_1)
\mathcal{M}_{2}(R_{\Sigma a \vec{a}},\vec{b}, -\vec{\lambda}_1) \frac{1}{4\pi} d\Omega_{\vec{\lambda}_1} = 
- \frac{ \vec{a} \cdot \vec{b}  }{4}.
\end{eqnarray}
Using \eqref{eq_evvt2}, it is easily seen that the HV distribution $R_{\Sigma a \vec{a}}$ of particle 2 depends both on the direction $\vec{a}$ and on the measurement outcome $a$ of the spin-measurement of particle 1. 
Both of these properties are in conflict with the definition of ``locality'' as given by Bell.

Matzkin does not give us any local mechanism (based on information locally available) to explain the perturbation of the HV distribution of particle 2 after measuring particle 1.  
So, we have to conclude that the \emph{initial} HV distribution $R_\Sigma$ of particle 2 is modified in a non-local way\footnote{
In fact, modifying a local variable (the HV-distribution) of particle 2 based on the know\-ledge we obtain by measuring particle 1, seems to us no more than some sophisticated way of formalizing the sentence:  ``by knowing the spin of particle 1 is $\suk{\vec{a}}$, measuring the spin of particle 2 in direction $\vec{b}$ should respect the quantum-mechanical predictions \emph{as if} particle 2 is in a quantum state $\sdk{\vec{a}}$''. This may be true, but can not be called a local model.
} to another HV distribution $R_{\Sigma a \vec{a}}$ when a measurement on particle 1  is made. 
Ironically, the only statement Matzkin made about locality is hereby not obeyed.

\section{Conclusion}

We recognize that measurement outcomes in the HV model of Matzkin not only depend on the HV $\vec{\lambda}$ and the the measurement direction $\vec{u}$, but also on a initial HV distribution $R$. 
In the HV model for the two particles in the EPRB experiment, the initial HV distribution of particle 2 is modified, based on both the measurement direction and the measurement outcome of a measurement on particle 1. 
Both of these properties of the HV model of Matzkin conflict with the criteria of a local theory, as defined by Bell.
Therefore we conclude that the given HV model is not local.

\end{document}